SUBMITTED FOR PUBLICATION IN PHYSICAL REVIEW B (JUNE 2005)

# Observation of Andreev bound states in $YBa_2Cu_3O_{7-x}$ /Au/Nb ramp-type Josephson junctions

B. Chesca [1*], D. Doenitz [1], T. Dahm [2], R. P. Huebener [1], D. Koelle[1], R. Kleiner [1], Ariando [3], H.J.H. Smilde [3], H. Hilgenkamp[3]

*1: Physikalisches Institut-Experimentalphysik II, Universität Tübingen, Auf der Morgenstelle 14, D-72076 Tübingen, Germany*
*2: Institut für Theoretische Physik, Universität Tübingen, Auf der Morgenstelle 14, D-72076 Tübingen, Germany*
*3: Faculty of Science and Technology and Mesa[+] Research Institute, University of Twente P.O. Box 217, 7500 AE Enschede, The Netherlands*

We report on Josephson and quasiparticle tunneling in $YBa_2Cu_3O_{7-x}$(YBCO)/Au/Nb ramp junctions of several geometries. Macroscopically, tunneling is studied in the *ab*-plane of YBCO either in the (100) and (010) direction, or in the (110) direction. These junctions have a stable and macroscopically well defined geometry. This allows systematic investigations of both quasiparticle and Josephson tunneling over a wide range of temperatures and magnetic fields. With Nb superconducting, the proximity gap induced in the Au layer appears in the quasiparticle conductance spectra as well defined coherence peaks and a dip at the center of a broadened zero-bias conductance peak (ZBCP). The voltage position of the coherence peaks varies with Au layer thickness. As we increase the temperature or an applied magnetic field both the coherence peaks and the dip get suppressed and the ZBCP fully develops, while states are conserved. With Nb in the normal state the ZBCP is observed up to about 77 K and is almost unaffected by an increasing field up to 7 T. The measurements are consistent with a convolution of density of states with broadened Andreev bound states formed at the YBCO/Au/Nb junction interfaces. Since junctions with different geometries are fabricated on the same substrate under the same conditions one expects to extract reliable tunneling information that is crystallographic direction sensitive. In high contrast to Josephson tunneling, however, the quasiparticle conductance spectra are crystallographic orientation *insensitive*: independent whether the tunneling occurs in the (100) or (110) directions, a pronounced ZBCP is always observed, consistent with microscopic roughness of the junction interfaces. Qualitatively, all these particularities regarding quasiparticle spectra hold regardless whether the YBCO thin film is twinned or untwinned. This suggests that the formation of Andreev bound states is, to a first approximation, insensitive to twinning.

## I. INTRODUCTION

Quasiparticle tunneling spectroscopy has been accepted to be one of the most sensitive probes of electronic states of superconductors. The appearance of a zero-bias conductance peak (ZBCP) in the differential conductance G = dI/dV versus voltage V, due to the formation of Andreev bound states (ABS) at interfaces involving $d_{x^2-y^2}$ (d)-wave superconductors is one of the most remarkable features distinct from conventional s-wave superconductors [1-7]. So far, ABS induced ZBCPs have been experimentally observed in three different systems: $NIS_d$, $S_dIS_d$, and $S_dIS_s$ junctions, where N is a normal metal, I is an insulator, and $S_d$ ($S_s$) is a d-wave (s-wave) superconductor. For $NIS_d$ junctions the formation of ABS and its implication on tunneling spectra has been intensively investigated experimentally and is well understood theoretically [5,6]. In contrast $S_dIS_d$ [8,9] and $S_dIS_s$ [10-12] junctions have been much less investigated experimentally. These systems are of particular interest since both Cooper pairs and quasiparticles are tunneling simultaneously. Therefore a direct comparison between the two tunneling channels from the point of view of potential tools for investigating the symmetry of the order parameter [2, 6, 7] is possible, although, as far as $S_dIS_s$ junctions are concerned all reports have concentrated on either quasiparticle tunneling [10-12] or Josephson tunneling [13]. In addition, as shown in the pioneering theoretical work [4], the energy gap of the $S_s$ superconductor appears on the conductance tunneling spectra of $S_dIS_s$ junctions in the form of a center dip. The observation of such a center dip superposed on a broader ZBCP rules out all the other alternative mechanisms (like supercurrent leakage, tunneling into a normal region, or scattering due to magnetic impurities in the barrier) that may induce a ZBCP as well, and therefore unambiguously proves the existence of ABS [4]. This is an important advantage $S_dIS_s$ have over $S_dIN$ and $S_dIS_d$ junctions that makes them very attractive tools to be used in phase-sensitive experiments to determine the symmetry of the superconducting order parameter. Finally, the striking similarity of the ZBCP in $NIS_d$ and $S_dIS_{d,s}$ ($S_{d,s}$ means that the superconductor is either a d-wave or an s-wave superconductor; $S_d$ = $YBa_2Cu_3O_{7-x}$ (YBCO) and $S_s$ = Nb in the case studied in this paper) systems is not well understood. There have been two scenarios proposed to explain such a similarity. Some authors consider that the ZBCP in the G(V) characteristic of $S_dIS_{d,s}$ junctions results from a convolution of density of states with strongly broadened mid-gap-states [4,8,9]. Other authors proposed a different model [6, page R72]. In their view for an $S_dIS_{d,s}$ junction to behave like a $NIS_d$ junction it is necessary that there are strong relaxation effects in the barrier region, the barrier acting as a reservoir, which results in decoupling of the two $S_dI$ and $IS_{d,s}$ interfaces. The resulting system will then be a series connection $S_dI+IS_{d,s}$ of two independent junctions (here we call this model [6] the $S_dI+IS_{d,s}$ model). For identical transparencies each junction would be biased by V/2 and the s-wave gap should occur well above $\Delta_s/e$. In contrast, in the frame of the $S_dIS_{d,s}$ convolution model the s-wave gap should occur at $\Delta_s/e$ and the d-wave gap appears at $\Delta_d/e$. Since there has been no independent measurement of the gap in experiments involving such junctions so far it remained an open question which of the models is appropriate for $S_dIS_{d,s}$ junctions.

* Corresponding author: boris.chesca@uni-tuebingen.de

In order to gain knowledge and a deeper understanding concerning all these issues related to tunneling from one superconductor to another, in this work we present measurements of both Josephson and quasiparticle tunneling in YBCO/Au/Nb ramp type junctions. As it will be clear later on, the quasiparticle conductance measurements suggest the presence of a thin insulating tunnel barrier which develops at the YBCO/Au interface, so that $S_dINS_s$ junctions are actually formed. Then, due to a proximity effect in the Au layer, such junctions behave like $S_dIS_s$ junctions. The junctions investigated here have the advantage of being realized in a well controlled junction geometry, so that macroscopically tunneling can be probed in different directions. This important feature allows systematic investigations of both quasiparticle and Josephson tunneling over a wide range of temperature and magnetic field for junctions of different orientations fabricated on the same chip. The Josephson channel may be switched off by a small applied magnetic field. That opens a unique opportunity to directly compare the two tunneling channels as tools to investigate the junction structure and the symmetry of the superconducting order parameter [2,6,7]. We will show that both tunneling methods are actually complementary to one another. In particular we present evidence of Andreev bound states formation at the YBCO/Au interface independent on the macroscopic tunneling direction, i.e., for both (100) and (110) junctions. On the other hand, the junctions investigated here offer the unique opportunity to clarify the issue which of the models, the $S_dIS_s$ convolution model or the $S_dI+IS_s$ model, is appropriate for tunneling from one superconductor to another. This is possible since the proximity gap induced by Nb in the Au layer is well defined and easy to interpret in the conductance spectra. Also the $S_d$ (YBCO) gap is a reasonably well known quantity from various types of experiments.

The paper is structured as follows: in section II we present a representative selection of the measurements performed: current –voltage characteristics (IVCs), Josephson critical current $I_c$ versus small applied magnetic fields B (in the $\mu$T range) and quasiparticle tunneling spectra G(V) for large applied magnetic fields (from 0.1 T up to 7 T) and variable temperature T (from 4.2 K up to 77 K), and for different Au barrier thickness. In section III we present theoretical approaches developed in an effort to explain both the Josephson and quasiparticle tunneling experimental data. Those calculations allow: a) to gain insights into the junction structure, in particular about the scattering rates at both YBCO/Au and Au/Nb interfaces and on the interface roughness, and b) to distinguish between the $S_dIS_d$ convolution model and the $S_dI+IS_{d,s}$ model.

## II. TUNNELING EXPERIMENTS IN YBCO/AU/NB JUNCTIONS

For the preparation of YBCO/Au/Nb ramp junctions (see Fig.1), bilayers of 150 nm [001]-oriented YBCO and 100 nm $SrTiO_3$ were grown by pulsed-laser deposition on edge-aligned [001]-oriented $SrTiO_3$ single crystal substrates. The structures were ion milled under an angle with the substrate plane yielding a ramp with a slope of $15^0$-$20^0$ [14]. Then, a 6 nm YBCO interlayer was prepared, providing for a high-quality interface to the subsequently *in situ* deposited Au-barrier layer. Finally, a 140 nm Nb layer was deposited by dc sputtering. Fabrication of these junctions [14] as well as Josephson tunneling in these systems [13] have been reported elsewhere. Measurements of the Josephson critical current versus a magnetic field applied either perpendicular or parallel to the *c*-axis, strongly suggests that electrical transport in these junctions occur essentially in the *ab* plane alone, so that the *c*-axis coupling may be neglected. For the experiments junctions of several geometries were used, see Fig.1. *Macroscopically,* tunneling occurs either in the (100) direction, or the (110) direction, or both (100) and (010) directions simultaneously. The directions are taken with respect to the YBCO crystallographic orientation. Following notations from [13] we call the latter type of junctions having N alternating (100) and (010) facets, zigzag junctions. We made four-point measurements in a liquid-helium cryostat (T = 4.2 K) to obtain the IVCs. Families of IVCs taken at various B values in an electrically shielded room and a magnetically well-shielded liquid-helium cryostat have been used to extract the $I_c(B)$ characteristics. The magnetic field was always applied along the (001) direction, i.e., perpendicular to the *ab* plane of YBCO. As far as quasiparticle spectra are concerned we numerically differentiate the IVCs to obtain the differential conductance. In total we measured 2 chips. The junctions fabricated on the two chips are characterized by a different Au barrier thickness: $d_{Au}$=12 nm for junctions on chip #1 and $d_{Au}$=30 nm for junctions on chip #2. Chip #1 contained 7 junctions with different geometries, and in this case the YBCO thin film was twinned. To make a selection, here we show measurements of a 50 μm wide (100) junction, a 280 μm wide (110) junction and 3 zigzag junctions: a junction with 10 facets of 40 μm (called 10 x 40 μm), a 40 x 5 μm junction, and a 8 x 25 μm junction. Since for the junctions on chip #1 the YBCO is twinned it means that in the case of a (100) or (010) facet there is always a combination of (100) and (010) tunneling directions. However, in the following, we will refer to (100) and (010) in the macroscopic tunneling directions simply to indicate a $90^0$ difference in the tunneling directions with respect to a fixed orientation in the *ab* plane of the substrate. As far as the junctions from chip #2 are concerned all junctions were 4 μm wide and we measured 4 of them for each of the orientation: (100), (010) and (110). It is also important to notice that in this case the YBCO thin film was untwinned. All of the experimental data presented except those from Fig.8 are for the chip #1 ($d_{Au}$=12 nm, twinned YBCO).

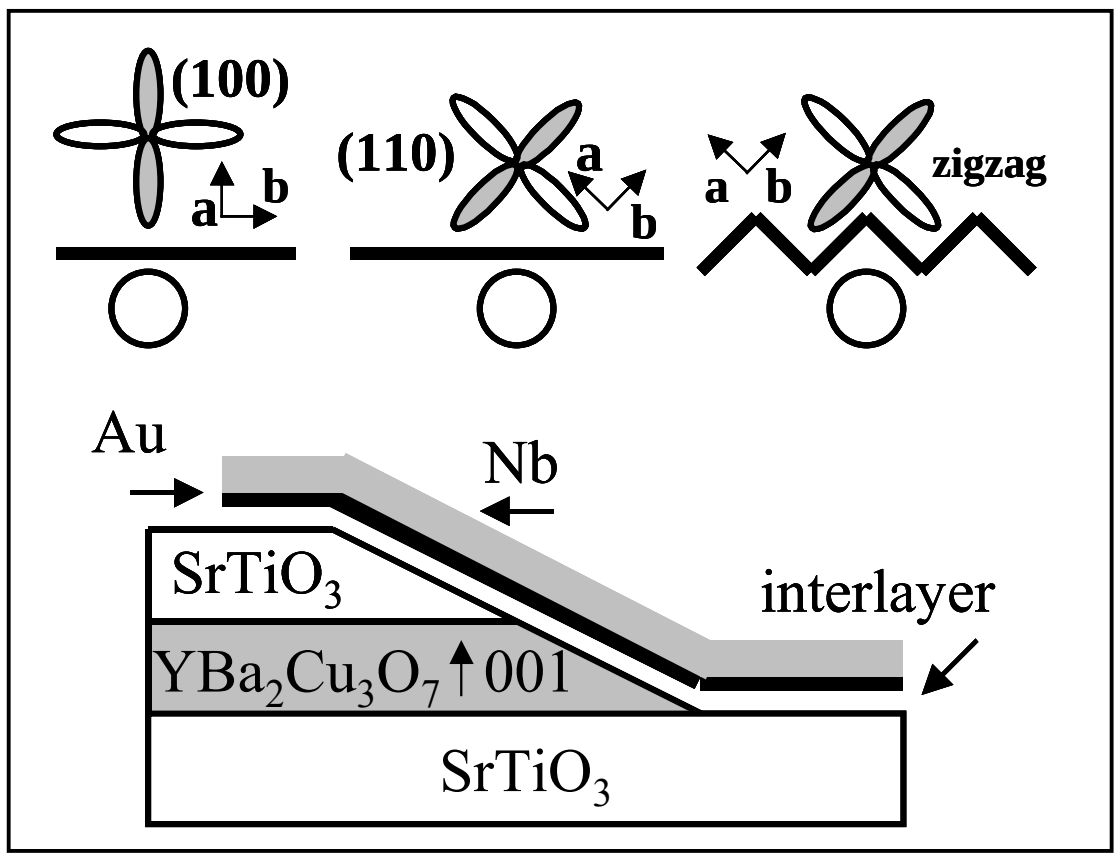

**FIGURE 1**: Schematic topview of a (100), a (110), and a zigzag junction and sideview of a $YBa_2Cu_3O_{7-x}$/Au/Nb ramp-type junction.

## A. JOSEPHSON TUNNELING

In accordance with earlier reports [13] for similar (100) and zigzag junctions, all junctions we measured at $T$ = 4.2 K and zero applied magnetic field have hysteretic IVCs (see the inset of Fig.2b), that are well described by the resistively and capacitively shunted-junction (RCSJ) model [15]. We measured $I_c(B)$ as a function of small applied fields B in the μT range. At fields in the mT range or higher there is no trace of a Josephson supercurrent left on the IVC. The (100) junctions have an $I_c(B)$ that qualitatively resembles a Fraunhofer pattern (see Fig.2(a); also [13]). For the (110) junction the $I_c(B)$ pattern (see Fig.2(b)) is *qualitatively* different and consists of many peaks with amplitudes randomly distributed, similar to the case of $45^0$ asymmetric bicrystal grain boundary junctions [16]. Such a pattern is clearly supportive of the *d*-wave symmetry of the superconducting order parameter in YBCO. It may be well understood qualitatively in terms of a junction interface that consists of a multitude of small junctions with positive and negative junction critical current densities [16,17]. For the zigzag junctions the $I_c(B)$ patterns (not shown here) are also anomalous with respect to the Fraunhofer pattern, similar to previous reports on similar junctions [13]. To conclude, Josephson tunneling is crystallographic orientation *sensitive*: there are sharp qualitative differences between various tunneling orientations.

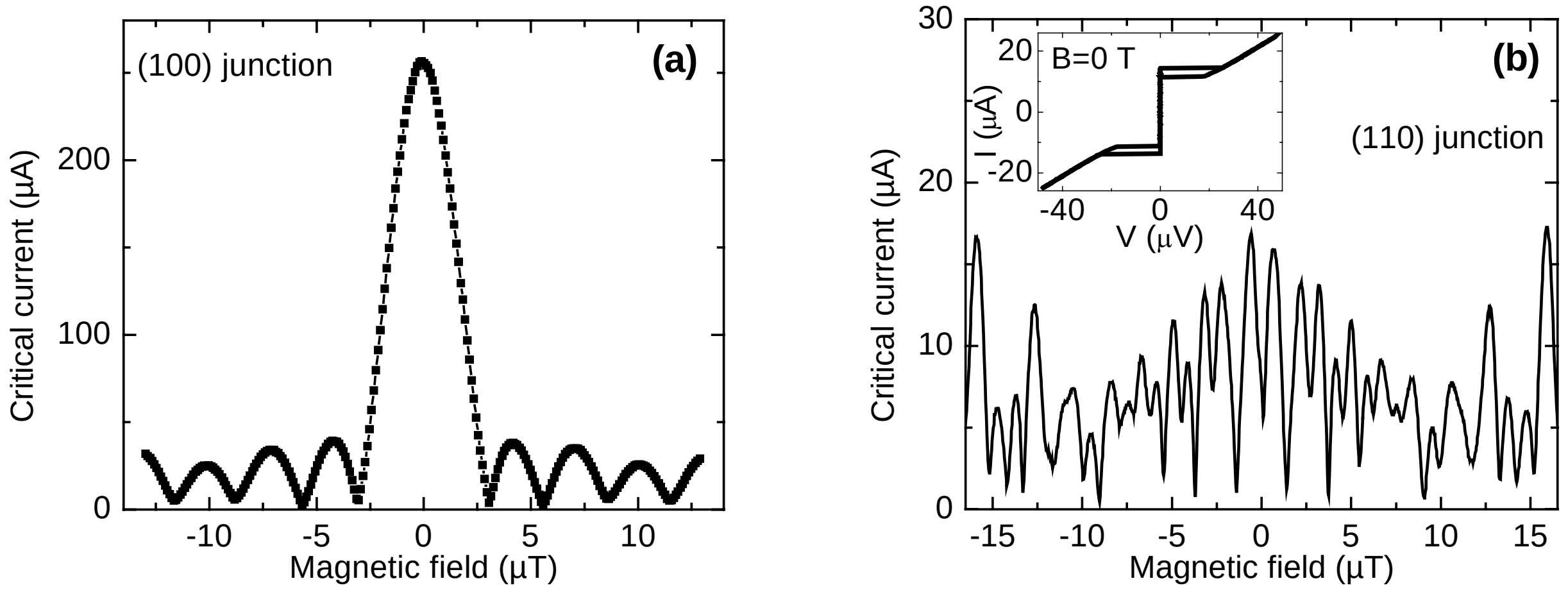

**FIGURE 2***:* Critical current versus magnetic field measured at 4.2 K for (a) the (100) junction, and (b) the (110) junction. Inset in (b) show current-voltage characteristics (IVC) at B=0.

## B. QUASIPARTICLE TUNNELING

Figure 3 shows representative quasiparticle tunneling data at B=0.01 T of the (100) junction, the (110) junction, and of two zigzag junctions (10 x 40 μm and 40 x 5 μm), at two different temperatures: 4.2 K and 9.05 K (just below the critical temperature of Nb, $T_{c,Nb} \approx 9.1$ K). At 4.2 K, with Nb superconducting, a proximity gap is induced in the Au layer that appears as pronounced gap-like features at about V = ±1.25 mV (for the 40 x 5 μm zigzag junction the coherence peaks are located at slightly larger voltages). For simplicity we call the observed coherence peaks Nb peaks, but we always have to keep in mind that those peaks are actually due to the proximity effect induced gap in the Au layer. As expected the coherence peaks occur at voltages close to the Nb gap values of ±1.4 mV as reported in YBCO/Nb tunnel junction measurements [10]. As calculations for quasiparticle tunneling show (see section III) the voltage position of the coherence peaks, however, is not exactly the value of the Nb induced proximity gap in the Au layer, but it is usually *larger* and it depends on the scattering rate in the YBCO layer within a coherence length from the YBCO/Au interface. At 9.05 K the Nb coherence peaks are completely suppressed and a pronounced ZBCP appears independent on the junction geometry. In some cases (see for instance the 10 x 40 μm zigzag junction shown again in Fig.4a for larger voltages) a clear gap like feature is observed at about ±19 mV that agrees well with other reported values in the literature for the YBCO gap [11,12].

We next discuss the B and T dependencies of the conductance spectra. Before we get into details, let us summarize some of the main results (cf. Figs. 4 and 5).

a) As we increase T from 4.2 K up to slightly below $T_{c,Nb}$, or B from 0.1 T up to slightly below the second critical field of Nb ($B_{c2,Nb} \approx 1.15$ T), we do observe one and the same qualitative picture which is *independent* of the crystallographic orientation. In other words, quasiparticle tunneling in (100), (110), or zigzag junctions look all alike.

b) As superconductivity gets suppressed in Nb, by increasing T or B the Nb coherence peaks get suppressed and the ZBCP gradually develops. Close to the critical values ($T_{c,\,Nb}$ or $B_{c2,\,Nb}$) of Nb there is no trace left of the Nb coherence peaks, while the ZBCP is fully developed.

c) Increasing T or B slightly above the critical values, there is a clear relatively sudden vertical shift of the conductance spectra to lower values of G(V). In addition to the superconducting-normal state transition of Nb, the ZBCP gets suppressed and it widens. These effects are due to the fact that now we measure the YBCO/I/Au junction resistance in series with the normal Nb layer. The degree of the sudden vertical shift of the conductance spectra as well as the degree to which the ZBCP gets suppressed and widens varies from junction to junction and increases with the magnitude of the Nb normal resistance relative to the junction YBCO/I/Au resistance. With Nb in the normal state, to get the voltage response of the junction YBCO/I/Au alone one has to extract the voltage drop due to the resistive Nb from the measured voltage. In doing so one recovers the amplitude and the width of the ZBCP measured with Nb very close to the normal state but still in the superconducting state.

d) Increasing T or B even further (from above $T_{c,\,Nb}$ up to 77 K, or from above $B_{c2,\,Nb}$ up to 7 T), however, a significant difference appears between the T and B dependence of G(V). The ZBCP (its amplitude and width) is essentially not affected by an increase of B, while by increasing T the ZBCP gets strongly suppressed and it widens. In particular, at 77 K we could not observe any trace of a ZBCP. In addition, by increasing T or B the conductance spectra are gradually shifted vertically to lower values due to the field and temperature dependence of the conductance of the Nb normal layer.

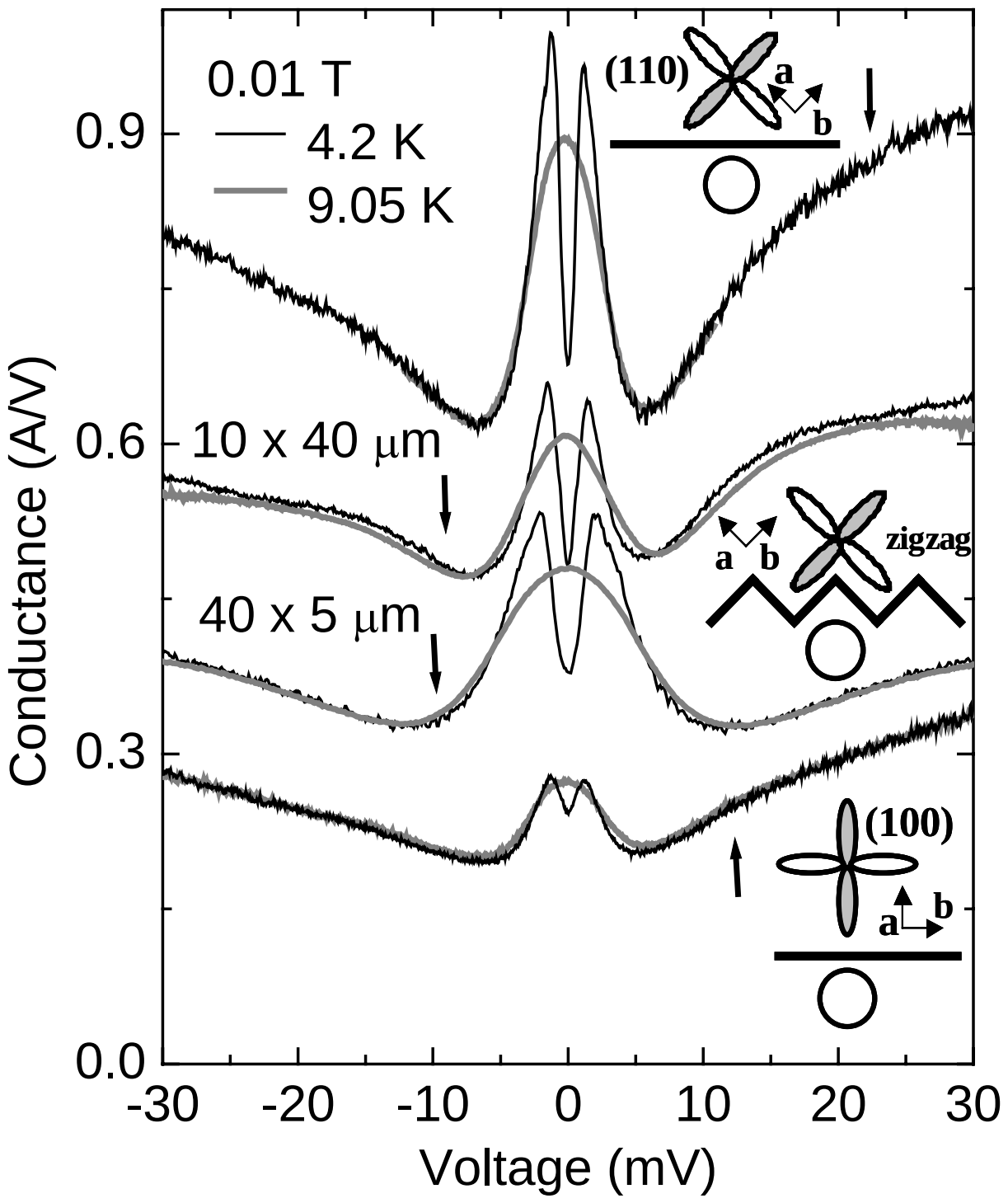

**FIGURE 3:** Tunneling spectra of 4 junctions having different geometries (see insets): the (100) and the (110) junctions, and the 10 x 40 μm and 40 x 5 μm zigzag junctions measured at two different temperatures: 4.2 K and 9.05 K (just below $T_{c,Nb}$). A magnetic field B=0.01T has been applied to completely suppress the Josephson critical current.

Figure 4(a) shows the B dependence of G(V) of a 10 x 40 μm zigzag junction from 0.01 T up to $B_{c2,Nb}$ at 4.2 K. In this case the YBCO coherence peaks are clearly seen at about ±19 mV. The Nb coherence peaks are gradually suppressed by an increasing field up to 1.05 T, as shown in the inset.

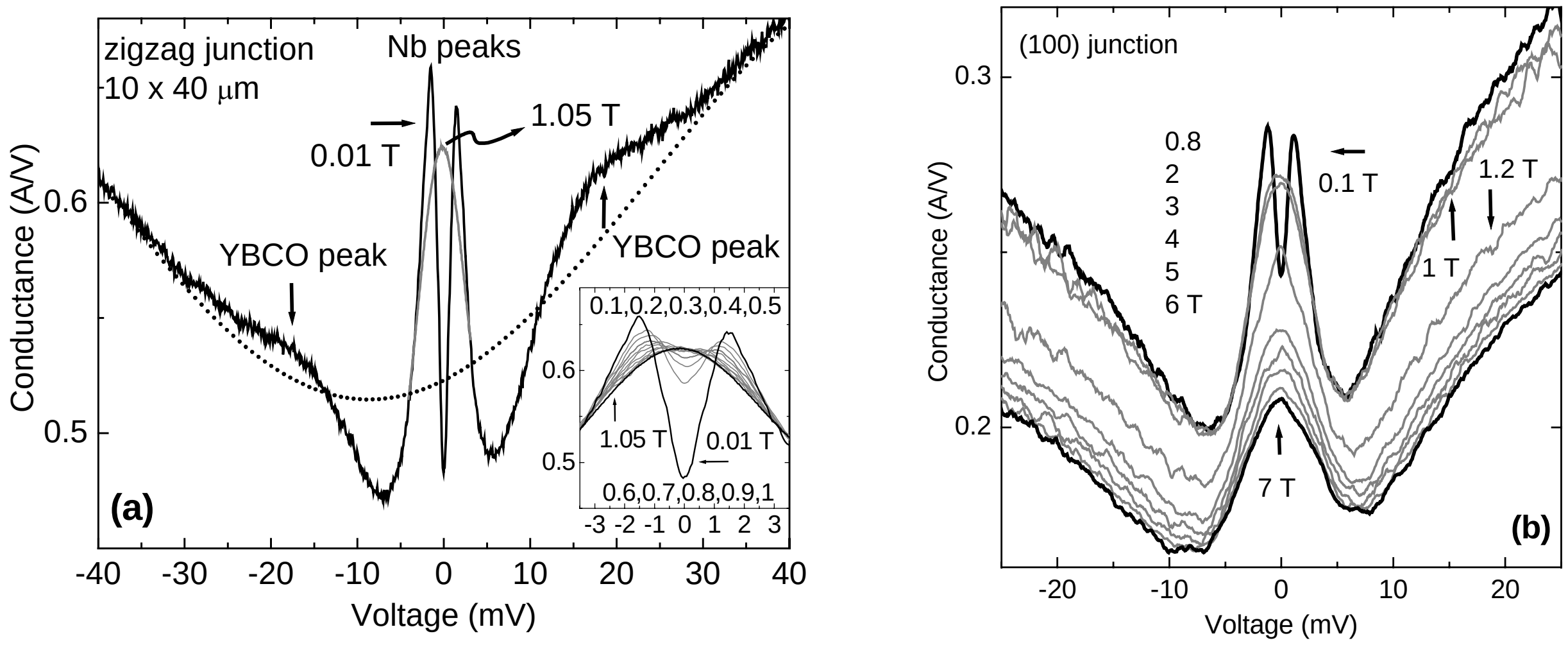

**FIGURE 4:** Tunneling spectra for different magnetic fields measured at 4.2 K. *(a)* 10 x 40 μm zigzag junction (shown also in Fig.3) for B = 0.01T – in black , and B = 1.05 T (just below $B_{c2,Nb}$) – in gray. Dotted line shows the background conductance. The inset shows the low voltage conductance spectra at 11 different field values. *(b)* (100) junction for ten different B values from 0.1 T up to 7 T;

Figure 4(b) shows 4.2 K conductance spectra of the (100) junction for 10 different values of B between 0.1 T and 7 T. Close to $B_{c2,Nb} \approx 1.15$ T, there is no trace left of the Nb coherence peaks, while the ZBCP is fully developed. As explained before, increasing B further, there is a significant sudden vertical shift of conductance at $B_{c2,Nb}$ (compare the curves for 1 T and 1.2 T). Increasing B further, over $B_{c2,Nb}$, up to 7 T the ZBCP is practically unaffected.

Figure 5 shows low voltage conductance spectra of a 40 x 5 μm zigzag junction for 10 different temperatures between 4.2 K and $T_{c,Nb}$. As superconductivity gets suppressed in Nb, by increasing temperature T the Nb coherence peaks get suppressed too and the ZBCP gradually develops. Figures 6a, 6b and 6c show the conductance spectra of the 8 x 25 μm zigzag junction for 11 different temperatures between 4.2 K and 77 K. For this particular junction the vertical shift of the spectra induced by the Nb transition from its superconducting state to its normal state (see Fig.6a) is more pronounced than e.g. for the (100) junction. In addition, at the transition the ZBCP widens and its amplitude gets suppressed. These changes of the ZBCP at the superconducting-normal state Nb transition are due to the fact that now we measure the YBCO/I/Au junction resistance in series with the normal Nb layer. To get the voltage response of the junction YBCO/I/Au alone one has to replace the voltages V measured at every current bias I by $V-IR_{Nb}$, with $R_{Nb}$ being the resistance at zero voltage. In doing so one recovers the amplitude and the width of the ZBCO measured with Nb very close to the normal state but still in the superconducting state (see Fig.6c). Increasing T further above $T_{c,Nb}$ the ZBCP is gradually suppressed and is not visible anymore at 77 K (see Fig.6c).

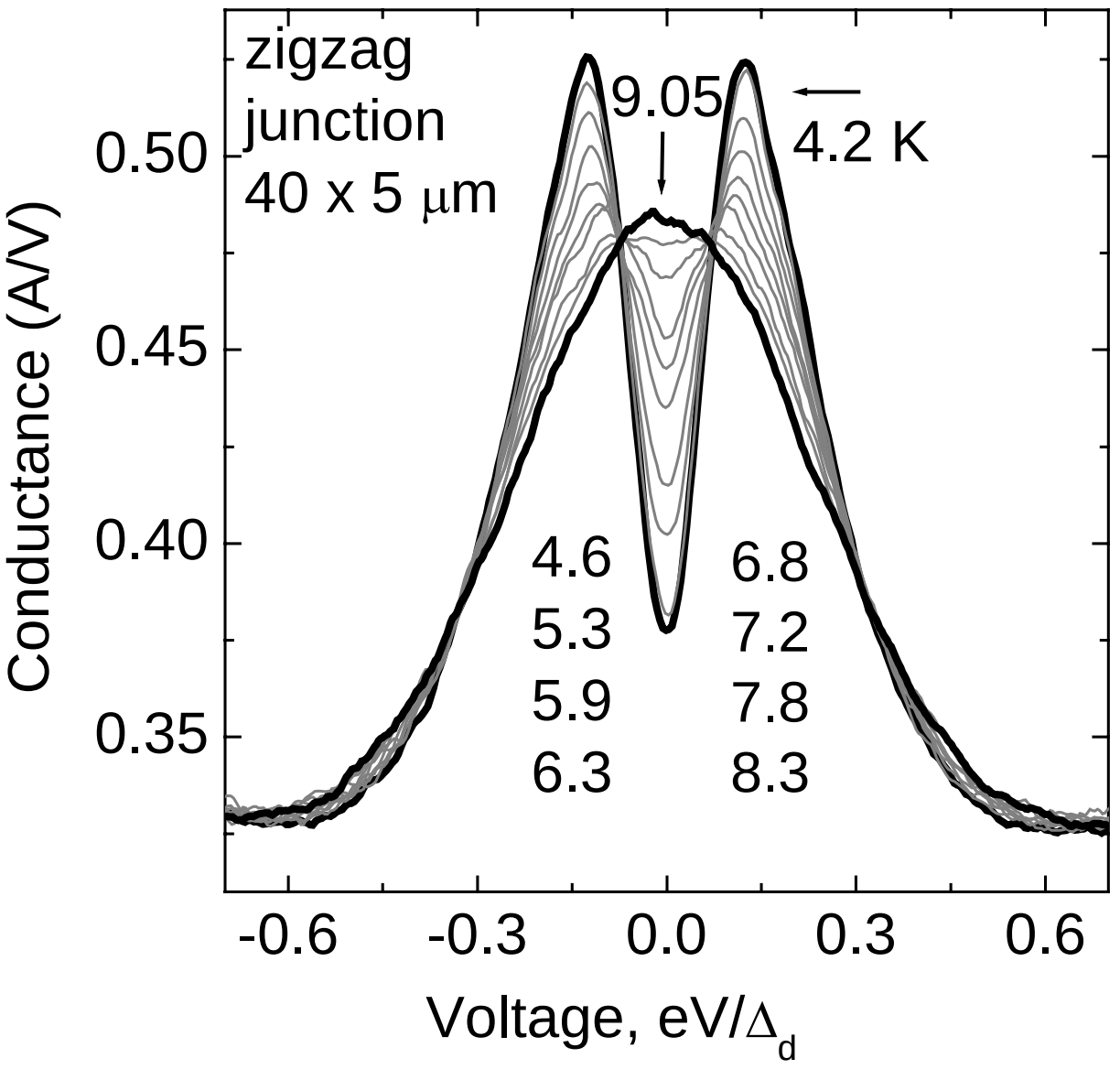

**FIGURE 5:** G(V) of a 40 x 10 μm zigzag junction at 10 different temperatures between 4.2 K and 9.05 K measured at B=0.01 T (just below $T_{c,Nb}$) with $\Delta_d$ =19 mV.

In both situations (by increasing either T or B) the total number of states between -3 mV and 3 mV is conserved to a remarkable degree, to within 95 %, supporting the interpretation of the conductance as a feature of a superconducting density of states. We have checked that precisely with similar results for many junctions by integrating the conductance spectra. One characteristic example is shown in Fig.7 for the zigzag junction 10 x 40 μm.

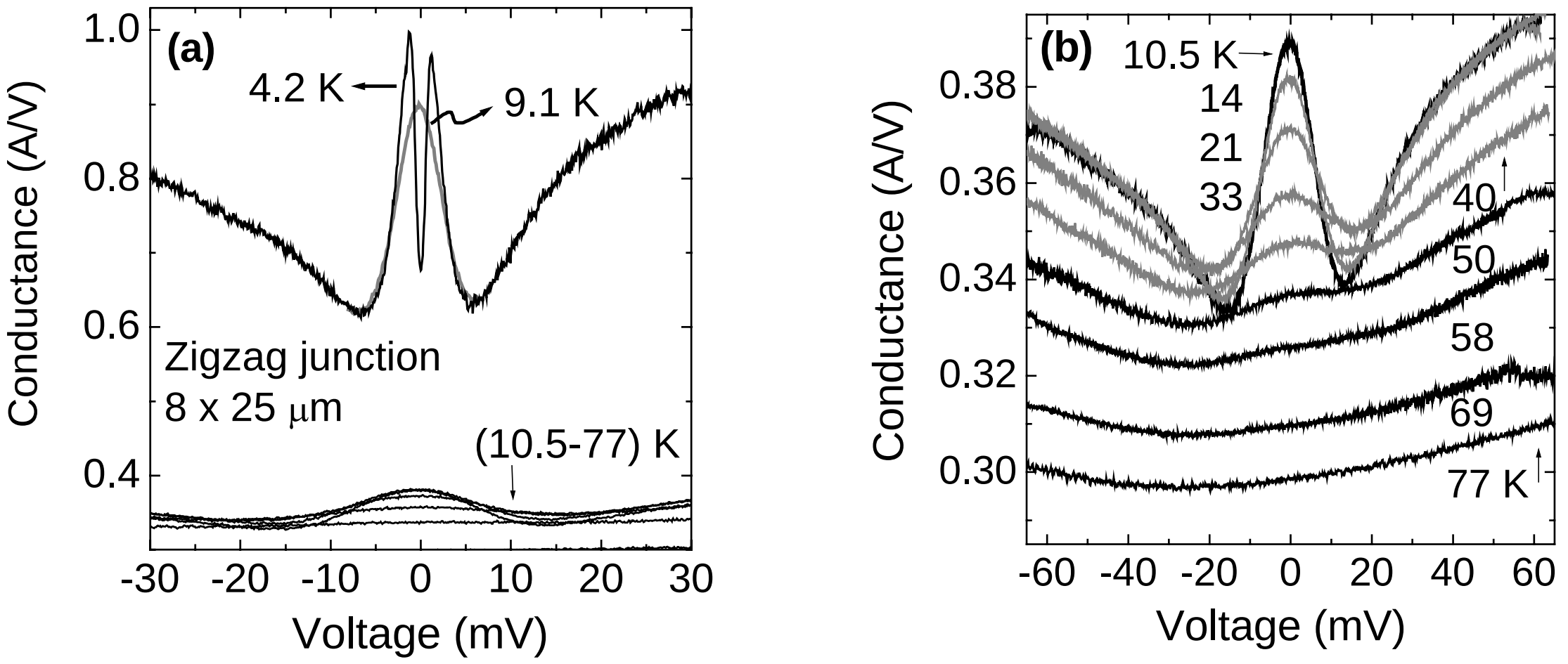

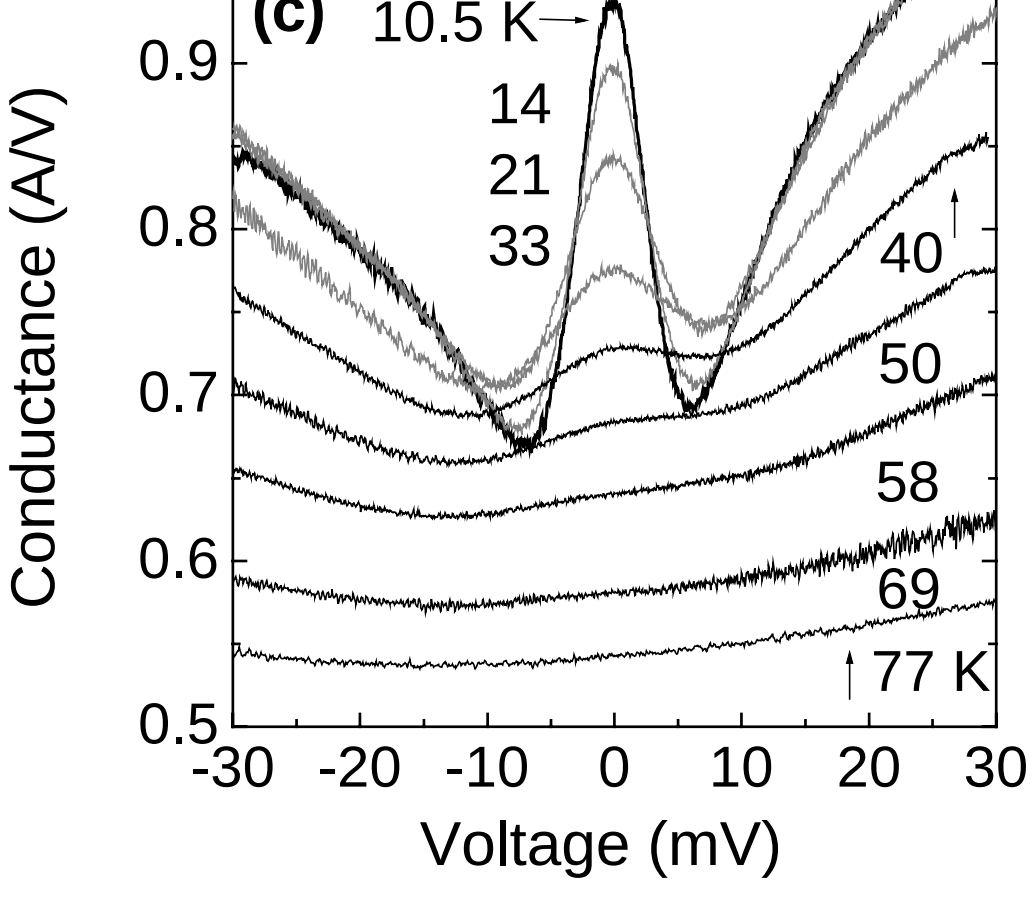

**FIGURE 6:** G(V) of a 8 x 25 μm zigzag junction measured at B=0.01 T and for 11 different temperatures between 4.2 K and 77 K. *(a)* G(V) at all temperatures. *(b)* G(V) above $T_{c,Nb}$ in detail; shown here is the total voltage response (as measured) of the YBCO/Au junction in series with the Nb normal state resistance, $R_{Nb}$. *(c)* Rescaled G(V) to show the YBCO/Au junction response alone, obtained from G(V) plotted in *(b)*. To obtain this graph we extracted the voltage drop due to the resistive Nb from the measured voltage V at each current bias I, i.e., we replaced V by V-$IR_{Nb}$, with $R_{Nb}$ being the resistance at zero voltage.

**FIGURE 7:** Normalized integrated density of states IDOS (between –3 mV and 3 mV) as a function of the normalized applied field $B/B_{c2,\ Nb}$ at T=4.2 K (black circles) and the normalized temperature $T/T_{c,\ Nb}$ at B=0.01 T (empty squares).

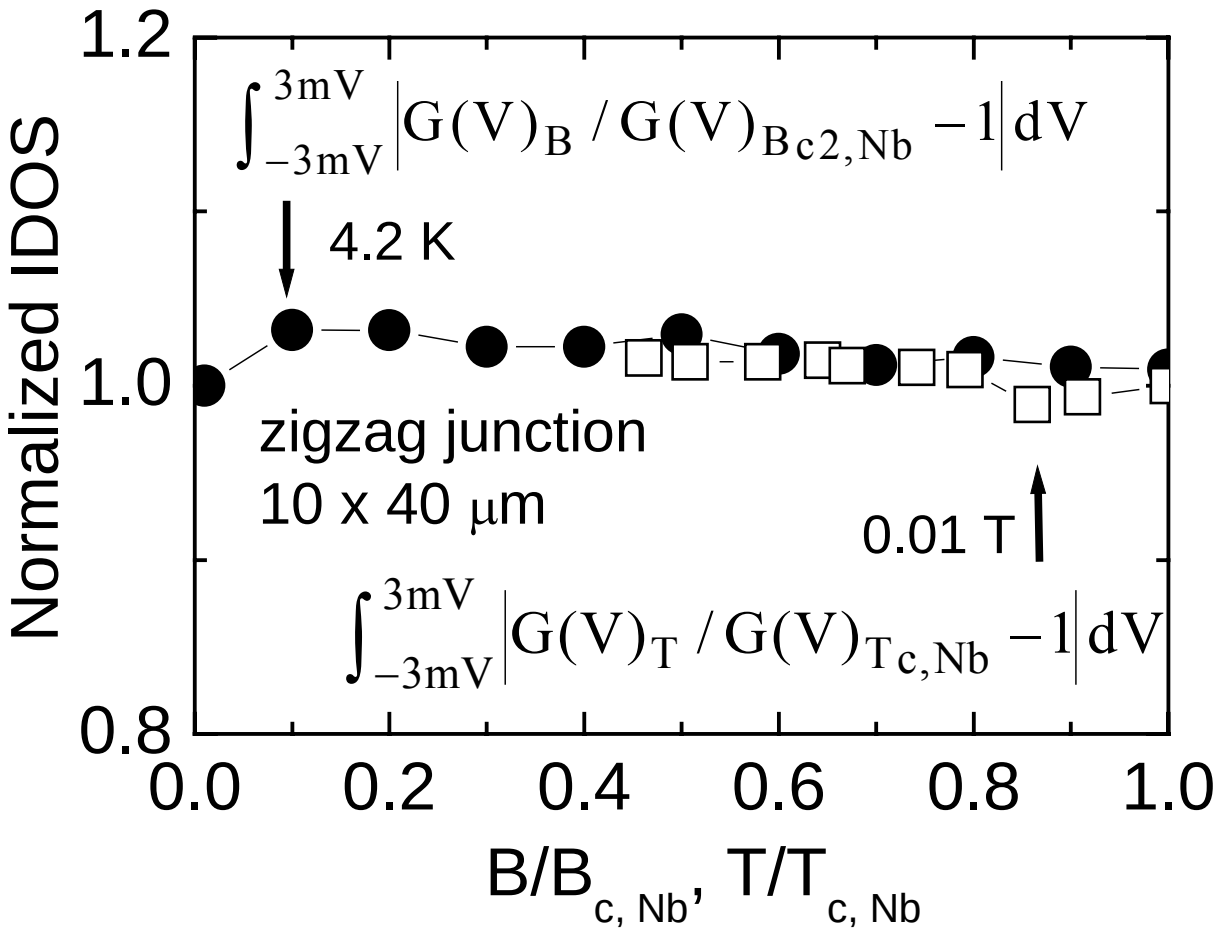

Apart from the tunneling spectra presented so far of junctions fabricated on chip #1 we also measured (100) and (110) junctions on chip #2 ($d_{Au}$=30 nm, untwinned YBCO thin film [19]). Josephson tunneling in junctions [20] made with an untwinned YBCO thin film have been reported elsewhere. The quasiparticle spectra showed no qualitative difference between junctions patterned on the two chips, which proves that the formation of Andreev bound states is not related to the twinned or untwinned character of the YBCO thin film. All features, (a) to (d), summarized in the beginning of Sec. II B remain valid for the junctions on chip #2 too. It means that, again, formation of Andreev bound states occurs for tunneling not only in the (110) direction but in the (100) direction as well (see the two coherence peaks and a dip at the center of a broadened ZBCP in Fig. 8). We found no qualitative difference between quasiparticle spectra of (100) and (010) junctions on chip #2 either, although, due to the untwinned character of YBCO thin film, macroscopically tunneling occurs *selectively* either in the *a* direction, or *b* direction of the YBCO film. Quantitatively, however, as far as junctions on chip #2 are concerned, there are only very small differences in the conductance spectra. On the one hand, this is what we expect since the junctions are all 4 μm wide, but on the other hand it proves that all junctions have very similar interface quality. A comparison between (100) and (110) junctions patterned on the two chips and measured at 4.2 K and B = 0.01 T is shown in Fig.8. The coherence peak position decreases with the Au thickness from about 1.25 mV for $d_{Au}$=12 nm to about 0.75 mV for $d_{Au}$=30 nm. That supports the interpretation that the measured gap is the proximity effect induced gap in the Au layer and not the Nb gap itself [18]. However, the mean free path of the electrons in the Au layer may change from one chip to another so that one should be careful in drawing conclusions about the proximity induced gap dependence on Au layer thickness from these data.

**FIGURE 8:** Low voltage tunneling spectra at T=4.2 K and B=0.01 T of (100) and (110) junctions patterned on two different chips (chip #1: $d_{Au}$=12 nm, twinned YBCO; and chip #2: $d_{Au}$=30 nm, untwinned YBCO). To facilitate the comparison we have multiplied by 5 the conductance of junctions from chip #2. The vertical arrows indicate the position of the Nb coherence peaks due to the proximity induced gap in Au.

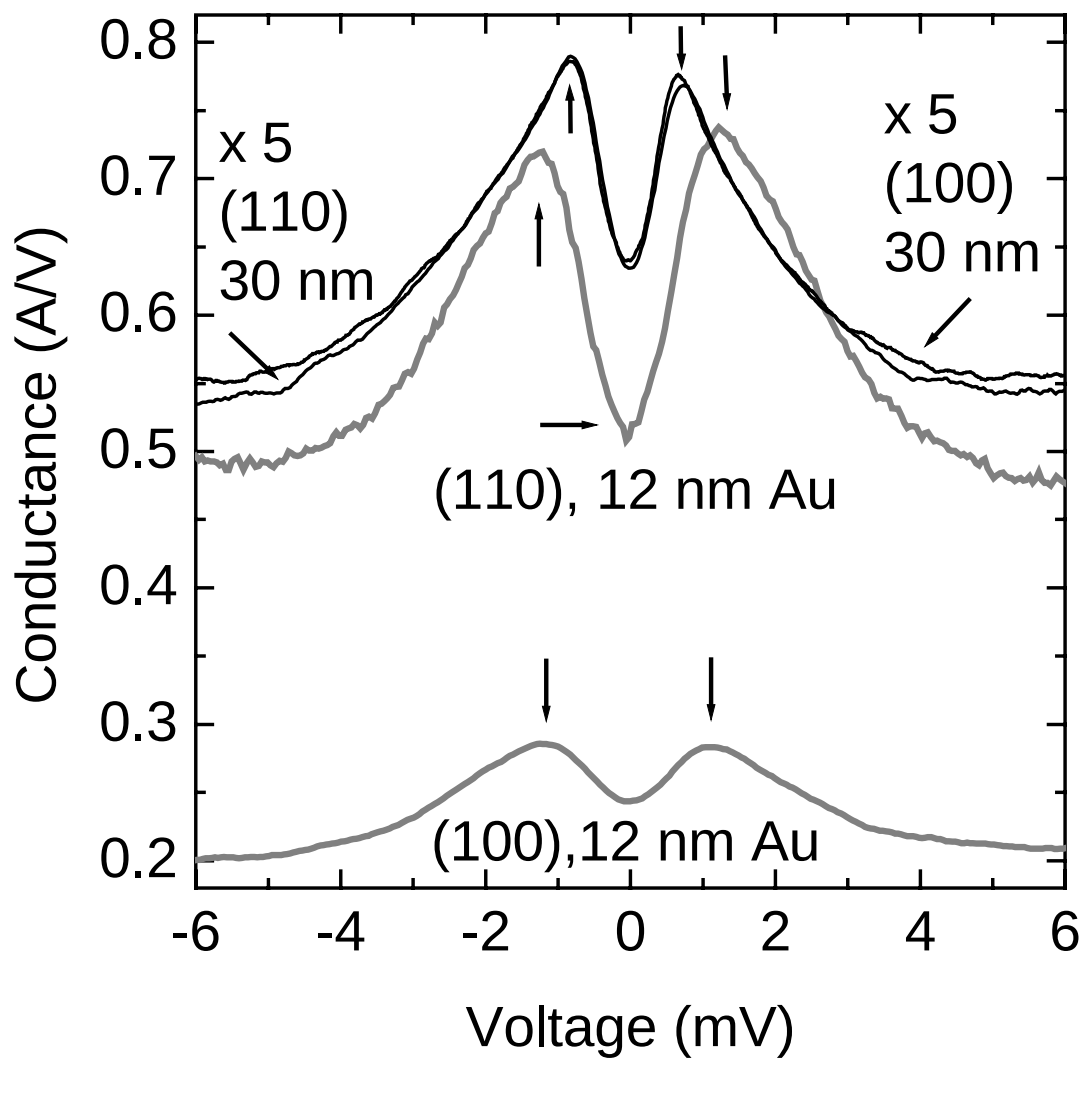

## III. MODELING OF JOSEPHSON AND QUASIPARTICLE TUNNELING.

In qualitative agreement with calculations made for $S_dIS_s$ tunneling junctions [4] the energy gap of the $S_s$ superconductor (Nb) appears in our G(V) measurements at 4.2 K in the form of a center dip (see Figs.3, 4, 5, 6 and 8) whose magnitude is determined by the spectral weight of the zero-energy states due to the *d*-wave character of the $S_d$ superconductor (YBCO). That indicates that tunneling largely contributes to the observed current in the junctions investigated here. It also suggests that the proximity effect induces superconductivity in the Au layer and consequently the junctions behave like $S_dIS_s$ junctions. However, there are important quantitative discrepancies between calculations from [4] and our measurements that we address here. In particular, in our case the ZBCP is much more pronounced, and the zero bias conductance (ZBC), i.e., G(V=0), below and above the critical temperature of the $S_s$ superconductor is comparable to the conductance at the gap voltage $\Delta_d \approx$19 mV of the $S_d$ superconductor, and thus much larger than G(V=0) calculated in [4].

To obtain a quantitative comparison with the measurements we have calculated the low transmission tunneling conductance of an $S_dIS_s$ junction in the absence of an applied field using quasiclassical techniques as described in [21,22]. In this case the normalized conductance is given by

$$\frac{G(V)}{G_n}=\frac{1}{G_n}\frac{dI}{dV}=\frac{d}{dV}\int_{-\infty}^{\infty} dE\; N^d(E)N^s(E+V)\left[f(E)-f(E+V)\right] \quad (1).$$

Here $f(E)=1/(1+\exp(E/k_BT))$ is the Fermi distribution function. $N^d(E)$ and $N^s(E)$ are the (normalized) local densities of states in the superconducting state on the *d*-wave and *s*-wave side, respectively. $G_n$ is the normal state conductance. On the *s*-wave side we take an *s*-wave density of states with a broadening parameter $\Gamma_s$ and a gap value $\Delta_s$ with a BCS temperature dependence:

$$N^s(E)=\mathrm{Re}\left[\frac{|E|+i\Gamma_s}{\sqrt{(|E|+i\Gamma_s)^2-\Delta_s^2}}\right]$$

How can one understand the crystallographic orientation *insensitivity* of quasiparticle spectra in the frame of the rough interface junction model? In other words, no matter whether tunneling occurs in the (100), (110) or both (100) and (010) directions simultaneously, with Nb superconducting the two coherence peak structure is always present, while with Nb in the normal state a pronounced ZBCP is always observed (see Fig.3). Indeed, for perfectly smooth interfaces no ZBCP is expected for (100) or (010) junctions, nor for zigzag junctions, while the ZBCP should reach its maximum in the case of (110) junctions. That strongly contradicts our observations. We argue here that it is the interface roughness that is responsible for this. Within the framework of the quasiparticle tunneling junction model used here, alternatively, one may call this “diffuse reflection” instead. In the case of $NIS_d$ junctions it has been shown [3, 21, 22] that even a weak interface roughness or/and faceting both induce strong similarities between conductance spectra of (100) and (110) junctions. To get a good agreement with the experimental observations we have extended the rough interface junction model to $S_dIS_s$ systems so that the calculated conductance spectra consequently will be crystallographic orientation *insensitive*. The roughness at the junction interface will smear out the uniqueness of the *macroscopic* tunneling orientation. Thus, on the *d*-wave side we average the surface local density of states over all possible angles φ of the surface with the orientation of the *d*-wave gap, in order to model the microscopic roughness of the junction. This gives the expression:

$$N^{d}(E) = \frac{1}{\pi^{2}} \int_{-\pi/2}^{\pi/2} d\varphi \int_{-\pi/2}^{\pi/2} d\theta \, \mathrm{Re}\left[ \frac{E_{+}E_{-} - \Delta_{+}\Delta_{-}}{E_{+}E_{-} + \Delta_{+}\Delta_{-}} \right]$$

where $\Delta_{\pm} = \Delta_{d}\sin(2\theta \pm 2\varphi)$ *and* $E_{\pm} = |E| + i\Gamma_{d} + \sqrt{(|E| + i\Gamma_{d})^{2} - \Delta_{\pm}^{2}}$. Here, $\Delta_d$ is the gap of the *d*-wave superconductor, $\Gamma_d$ is a broadening parameter for the *d*-wave superconductor, and θ is the incident quasiparticle angle to the interface normal. The angle φ determines the orientation of the surface relative to the orientation of the *d*-wave gap. Some results for the tunneling conductance given by Eq. (1) are shown in Figs. 9(a) and 9(b) for $T/T_{c,s}$=0.5 and 1, and different values of $\Gamma_d$. Here, $T_{c,s}$ is

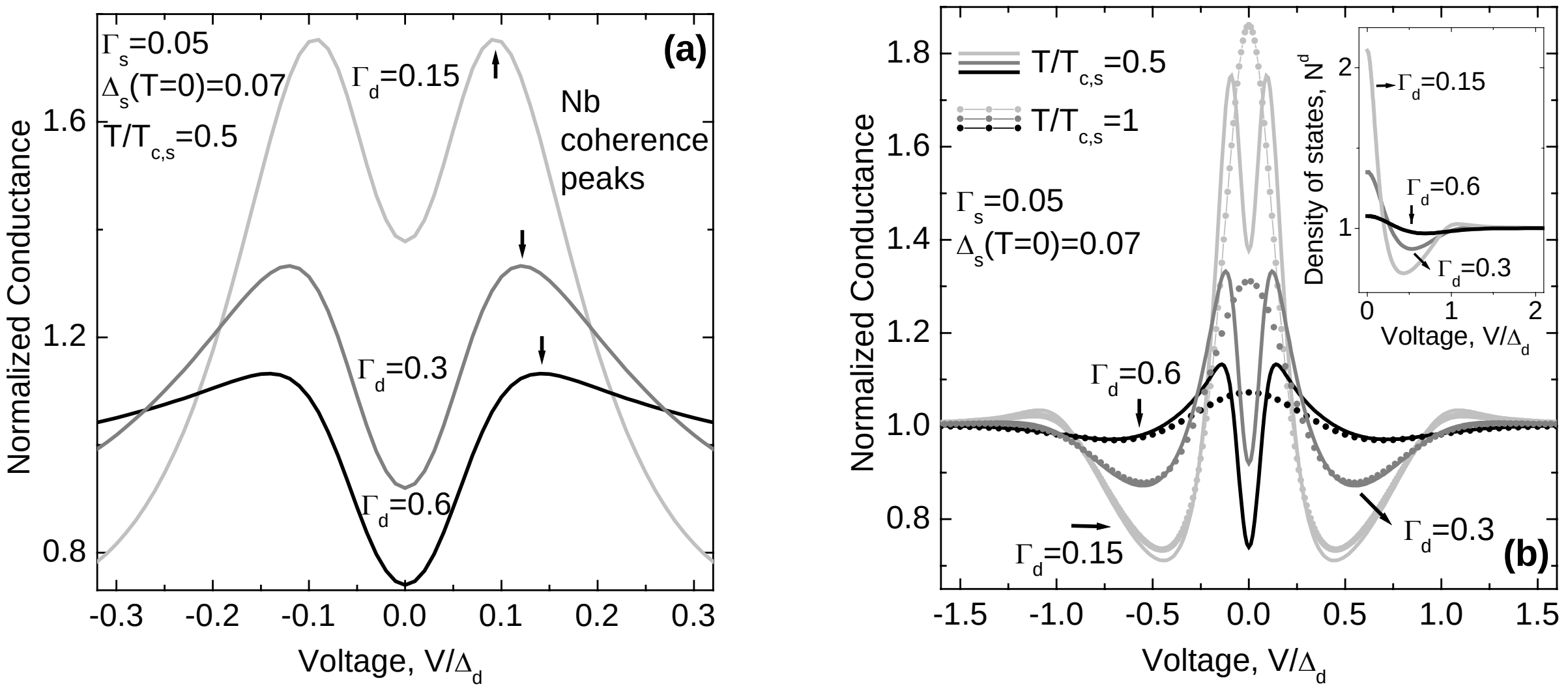

**FIGURE 9***:* Calculated normalized $G(V)/G_N(V)$ spectra in the frame of the $S_dIS_s$ model. a) Low voltage spectra for one temperature and three different values of the zero energy quasiparticle damping rate $\Gamma_d$ at the YBCO *d*-wave superconductor interface. b) The same as (a) but for two temperatures and for larger voltages. The inset shows the density of states in the *d*-wave superconductor (YBCO). $\Gamma_d$, $\Gamma_s$ , $\Delta_s$ values are normalized to $\Delta_d$=1.

the critical temperature of the *s*-wave side. In accordance with the measurements (see Figs.3, 4, 5, 6, and 8) the *s*-wave gap appears in the quasiparticle conductance spectra as well defined coherence peaks and a dip at the center of a broadened zero-bias conductance peak, independent on the tunneling direction. From the low voltage calculated spectra at $T/T_{c,s}$=0.5 shown on Fig. 9 (a) it follows that the coherence gaps are located as voltages larger than the s-wave superconductor gap value of 0.07 (for example at 0.09 for $\Gamma_d/\Delta_d$ = 0.15 and at about 0.14 for $\Gamma_d/\Delta_d$=0.6). At $T/T_{c,s}$=1 the coherence peaks are fully suppressed and the ZBCP fully develops (see Fig. 9 (b)). In this case one reaches the behavior of a $S_dIN$ junction instead, in accordance with the inset of Fig. 9(b) where we show $N^d(E)$ for different values of $\Gamma_d/\Delta_d$ = 0.15, 0.3 and 0.6. We have chosen $\Delta_s$=0.07$\Delta_d$ and $\Gamma_s/\Delta_d$ = 0.05, which gives the best fits to the experimentally observed *s*-wave peak structures for junctions on chip #1. The ZBCP is strongly suppressed, and the ZBC strongly decreases with increasing broadening parameter $\Gamma_d$ (see Figs. 9 (a) and 9 (b)). On the other hand, it is well known [23] that the ZBC monotonically increases with increasing $\Gamma_s$. Therefore the non-zero values for $\Gamma_s$ naturally explains, in particular, why in our case, as well as, in [10] the ZBC, when the *s*-wave side is superconducting, is considerably larger than observed in other systems [11-12] and as predicted in [4].

On the other hand, how can we understand the $I_c(B)$ measurements that are strongly crystallographic orientation *sensitive* in the frame of the rough interface junction model? To answer this question we have calculated the $I_c(B)$ pattern for both (110) (Fig. 10a) and (100) junctions (Fig.10b) according to the formula [24]:

$$I_c(B) = \int_{-\infty}^{+\infty} \left| J_c(x) e^{2\pi d i x B/\Phi_0} \right| dx \qquad (2).$$

Here d = $\lambda_{Nb}+\lambda_{YBCO}$+t, with t being the physical barrier thickness, and $\lambda_{Nb}$ ($\lambda_{YBCO}$) the London penetration depth in Nb (YBCO). From Eq. (2) it follows that $I_c(B)$ is the modulus of the Fourier transform of the critical current density $J_c(x)$ profile along the junction width L, with $0<x<L$. That means very accurate solutions for $J_c(x)$ at very small length scale (nm range) may only be tested for $I_c(B)$ measurements performed up to infinitely large magnetic fields. This is an impossible task in practice. Our primary goal therefore is not to obtain a very good quantitative agreement with the measurements but to show that a model of rough junctions, that comes out from quasiparticle tunneling, is consistent with the Josephson measurements as well (see Fig.2). In our model the tunneling orientation deviates locally from the macroscopic one and consequently is a function of coordinate *x*. Assuming YBCO has a *d*-wave symmetry of the order parameter it means $J_c$ is a function of *x,* too. We therefore model the junction roughness as a *continuous* variation of $J_c(x)$ [25]. The tunneling direction deviations induce a significant change in the amplitude of $J_c(x)$ only on a scale that is about 2 orders of magnitude smaller than the junction widths. We do not exclude that significant changes of $J_c(x)$ may actually occur at a smaller scale too, but the signature of that would not appear on the $I_c(B)$ measurements we performed at small fields. For consistency, we assume a *similar x*-dependent $J_c(x)$ profile for both (100) and (110) junctions (see insets of Figs.10(a) and 10(b)). Thus, for the (110) junction $J_c(x)$ oscillates randomly between -0.7 and 0.82 (in arbitrary units) around the average value $\langle J_c(x)\rangle = 0.06$. For the (100) junction it oscillates *alike* between 60 and 80 units around the average value $\langle J_c(x)\rangle = 70$. The difference in $\langle J_c(x)\rangle$ between (110) and (100) junctions is due to the *d*-wave symmetry model of the order parameter in which $J_c$ is a strongly angle dependent function and is expected to be zero if tunneling occurs *exactly* into the (110) direction and to have a certain maximum value for tunneling into the (100) direction. A reasonable qualitative agreement with the experiments (see Fig.2) is found as far as the essential features of the $I_c(B)$ patterns are concerned. For the (100) junction (see Fig.10(b)) one has a strong central maximum at B=0, followed by symmetrically distributed significantly smaller peaks. For the (110) junction one has a non-zero minimum at B=0, followed by a series of symmetrically distributed maxima characterized by amplitudes that differ randomly from one another. The huge difference in the maximum $I_c(B)$ values between the (100) and the (110) junctions (observed in the experiments, and confirmed by our calculations) can therefore be understood in terms of the (110) junction consisting of alternating regions with positive and negative junction critical current densities (see inset of Fig.10(a)). We simulated many other $J_c(x)$ profiles for both (100) and (110) junctions and compared them with the measurements. Here are some of the conclusions. Josephson tunneling measurements may provide a characteristic scale for the junction roughness. We found an upper limit of about a few hundreds of nm for significant non-uniformities in $J_c(x)$ of (100) junctions that are induced by the junction roughness. For (110) junctions we found that there must be significant $J_c(x)$ non-uniformities at a scale of about 10 μm. In addition, we cannot rule out that there might be significant $J_c(x)$ non-uniformities at a much smaller scale (nm scale), too. Moreover, this is most probably the case, since we needed to average over all possible tunneling directions (from $0^0$ up to $180^0$) in the quasiparticle calculations in order to find a good agreement with the measurements.

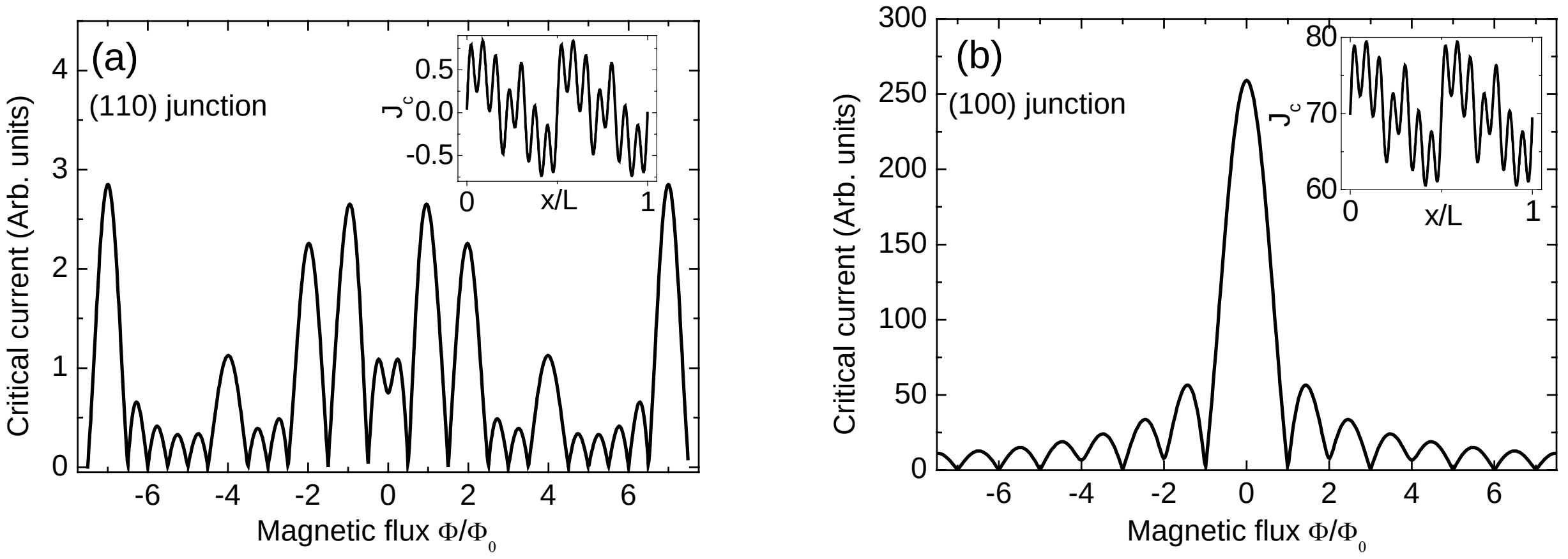

**FIGURE 10***:* $I_c(B)$ patterns calculated with Eq.(2) for *(a)* a (100) and *(b)* a (110) Josephson junction formed between *d*-wave superconductors. We assume a $J_c$ profile along the junction width L as shown in the insets. For the (100) junction the $J_c$ profile is similar to the case of the (110) junction but multiplied by 10 and shifted vertically by 16.5 so that $J_c$ is always positive.

Another interesting question we are able to answer is which of the two models for tunneling from a *d*-wave superconductor into an *s*-wave one is appropriate in our case: the $S_dIS_s$ convolution model or the series connection $S_dI$+$IS_s$ model [6, page R72] ? A significant difference is found between the two models in the location of the $S_d$ and $S_s$ coherence peaks which allows us a direct comparison with the measurements. The coherence peaks in the measured conductance spectra occur (see Figs.3, 4, 5, 6 and 8) at about the gap voltages of $S_d$ (of about ±19 mV for YBCO [11-12]) and $S_s$ (the proximity effect induced gap in the Au layer of slightly below the Nb gap of ±1.4 mV) in the $S_dIS_s$ model. On the other hand, for the $S_dI$+$IS_s$ model the coherence peaks occur essentially at *larger* values to a degree that depends on the relative conductances of $S_dI$ and $IS_s$ interfaces. For instance, for equal conductances of the two interfaces the coherence peaks occur at about double those values [6], while for the $S_dI$ interface having much smaller conductance than the $IS_s$ interface, the Nb coherence peak occur at a much larger value than ±1.4 mV. In the experiments, however, we have clearly observed coherence peaks located at about ±19 mV (for some junctions the YBCO peaks are less evident like in Figs.6a, 6b and 6c, for others they are better pronounced –see Fig.4a) and about ±1.25 mV for junctions with $d_{Au}$=12 nm (or ±0.75 mV for junctions with $d_{Au}$=30 nm) in all of the junctions measured which strongly suggests that the $S_dIS_s$ convolution model is appropriate in our case.

To perform a quantitative comparison of measured spectra with theory the experimental data have to be normalized to the normal-state background conductance. Experimentally we first have to increase temperature or magnetic field just over their critical values $T_c$ and $B_{c2}$ to suppress the superconductivity in the $S_d$ superconductor, and then measure the normal-state background conductance. The problem is that $B_{c2}$ of YBCO is difficult to reach in the experiments, while the normal-state background conductance may well be temperature dependent [2,6,12]. This is a serious complication that may distort the data and introduce significant errors [12]. To avoid these difficulties we normalize the measured spectra shown in Fig.5 to the curve at the critical temperature of Nb instead, as shown in Fig.11(a). Theoretically we first vary $\Gamma_d$, until the best fit is found with the measured G(V) at 9.05 K (the critical temperature of Nb, $T/T_c$ =1). Then we calculate G(V) at various temperatures $T/T_c$ <1, normalize them by the calculated curve at $T/T_c$ =1, and finally compare them with the experiments. The agreement between Figs. 11(a) and 11(b) is remarkable. This shows that one is able to characterize the quasiparticle scattering rates at both interfaces (YBCO/Au and Au/Nb) and learn about the interface nature.

## IV. CONCLUSIONS

In summary we measured temperature, magnetic field, and crystallographic orientation dependencies of the quasiparticle tunneling spectra of YBCO/I/Au/Nb junctions. As superconductivity gets suppressed in Nb by increasing either temperature or magnetic field the Nb coherence peaks get suppressed and the ZBCP gradually develops, while the total number of states is conserved. The measurements are consistent with the formation of Andreev bound states at the YBCO/Au/Nb junction interfaces and support a *d*-wave symmetry of the superconducting order parameter in YBCO. Conductance measurements of these junctions offered the unique possibility to test both earlier proposed models of Andreev bound states assisted quasiparticle tunneling from one superconductor to another: the $S_dIS_s$ convolution model and the series connection $S_dI$ + $IS_s$ of two decoupled interfaces model, and to prove the validity of the first one in our case. In high contrast to Josephson tunneling, the conductance spectra are crystallographic orientation *insensitive*: independent whether the tunneling occurs in the (100), (110) or both (100) and (010) directions simultaneously, a pronounced ZBCP is always observed. It follows that in our case, while both Josephson tunneling and quasiparticle tunneling are able to distinguish between *s*-wave and a superconducting order parameter with nodes, only Josephson tunneling is able to distinguish between *d*-wave and other possible symmetries characterized by a sign change, like $d_{xy}$-wave, etc. This is of importance when investigating the symmetry of the order parameter in a new, presumably unconventional superconductor. The crystallographic orientation insensitivity of the quasiparticle spectra proves that the junction interface is rough. Alternatively one may conclude that a diffuse reflection occurs at the junctions interfaces. Qualitatively, all these particularities regarding quasiparticle spectra hold regardless whether the YBCO thin film is twinned or untwinned. That proves that the formation of Andreev bound states is, to a first approximation, insensitive to the twinned or untwinned character of the YBCO thin film. On the other hand Josephson tunneling suggests that (110) junctions consist of alternating regions with positive and negative critical current densities. These differences show that Cooper pair and quasiparticle tunneling are complementary tools of investigation.

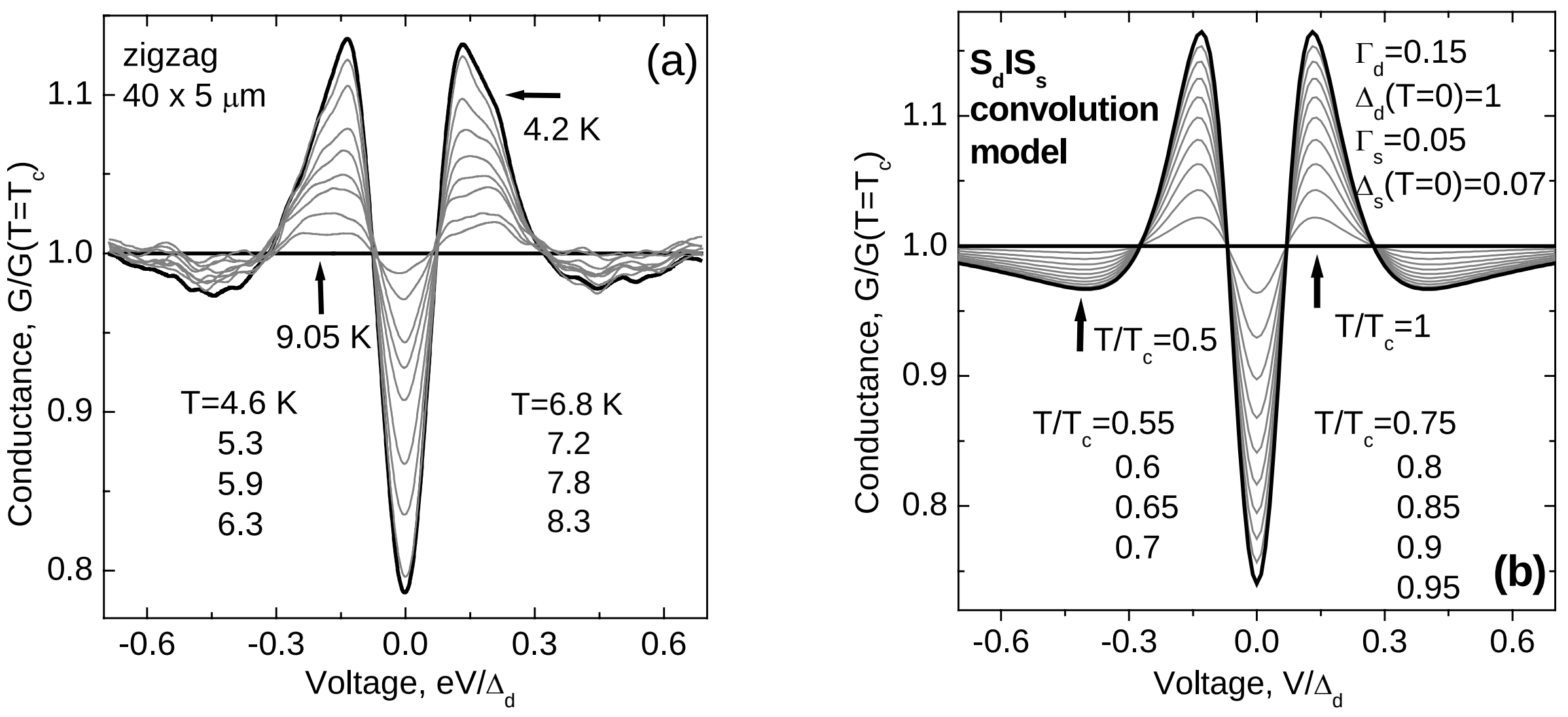

**FIGURE 11***:* Normalized conductance spectra (a) of the 40 x 5 μm zigzag junction (shown also in Fig.5) measured at 10 different temperatures between 4.2 K and 9.05 K with $\Delta_d$ =19 mV . (b) Calculations of the tunneling spectra versus temperatures performed with Eq.(1) in the frame of the the $S_dIS_s$ convolution model. $\Gamma_d$, $\Gamma_s$ , $\Delta_s$ values are normalized to $\Delta_d$=1.

## ACKNOWLEDGMENTS

Discussions with K. Scharnberg and A. A. Golubov are gratefully acknowledged. This work was supported by the ESF program PiShift, the Landesforschungsschwerpunktsprogramm Baden-Württemberg and the Dutch FOM and NWO foundations. D. Doenitz acknowledges support from the Evangelisches Studienwerk e.V. Villigst.